\documentclass[aps,pra,twocolumn,showpacs,floatfix]{revtex4}
\usepackage{epsfig}
\usepackage{graphicx}
\usepackage{dcolumn}
\usepackage{longtable}
\usepackage{amsthm,amsmath}

\begin{document}

\title{Development of a relativistic coupled-cluster method for one electron detachment theory: Application to Mn IX, Fe X, Co XI and Ni XII ions}
\vspace{0.5cm}

\author{D. K. Nandy \footnote{Email: dillip@prl.res.in} and B. K. Sahoo \footnote{Email: bijaya@prl.res.in}}
\affiliation{Theoretical Physics Division, Physical Research Laboratory, Ahmedabad-380009, India}

\date{Received date; Accepted date}

\vskip1.0cm

\begin{abstract}
We have developed one electron detachment theory from a closed-shell atomic configuration
in the relativistic Fock-space coupled-cluster ansatz. Using this method, we determine 
sensitivity coefficients to the variation of the fine structure constant in the first three important
low-lying transitions of the astrophysically interesting highly charged Mn IX, Fe X, Co XI and 
Ni XII ions. The potential of this method has been assessed by evaluating the detachment energies of the
removed electrons and determining lifetimes of the atomic states in the above ions. To account the 
sensitivity of the higher order relativistic effects, we have used the four component wave functions
of the Dirac-Coulomb-Breit Hamiltonian with the leading order quantum electrodynamics (QED) corrections.
A systematic study has been carried out to highlight the importance of the Breit and QED interactions
in the considered properties of the above ions. 
\end{abstract} 

\maketitle

\section{Introduction}
New field of research in the investigation of temporal variation of the fundamental constants has been gaining
the ground steadily in both the theoretical and experimental physics for the past few years \cite{Uzan,Budker, 
King, Chand}. The important aspects
for searching this variation are to establish theories suggesting violation of the Einstein's equivalence principle
and to support the models like Kaluza-Klein theory that attempts to unify gravity with the other three unified fundamental 
interactions \cite{Damour1, Damour2, Kaluza, Klein, Chodos, Marciano, Bronnikov}. This may also probe the multi-dimensionality 
to space as predicted by the superstring theories \cite{Damour3}. These theories predict temporal variation of the 
fundamental constants including the electromagnetic fine structure constant ($\alpha_e= \frac{e^2} {\hbar c}$) in the
low energy limit at the cosmological time-scale.

From the experimental front, the signature of possible variation of $\alpha_e$ can be observed
from three classes of measurements. {\it Geophysical method}: The data obtained from the isotopic decay
in the natural radioactive 
reactor at the Oklo observatory can be used as a tool to probe the variation of $\alpha_e$ as these
isotopes corresponds to typically $10^9$ years old and signature of the discrepancy between the decay rate
with the present laboratory value indicates a small deviation in the $\alpha_e$ value in this time scale
\cite{Uzan}. {\it Atomic clock method}: Also, the high precision frequency measurements using the atomic and 
the singly charged ionic clocks can be used to probe variation of $\alpha_e$ in an elegant manner. In 
these experiments the time dependency of $\alpha_e$ is inferred by comparing the transition frequencies 
between at least two clocks \cite{Uzan, Budker, Kolachevskil, Prestage}. The main advantage of these types of 
experiments is related to their efficient control over the systematic errors. However, the time period 
involved to carry out these measurements is of the order of few years only and the typical choice of the candidates 
for the clocks are either the neutral atoms or the singly charged ions where the relativistic enhancements are 
typically small. {\it Astrophysical method}: The most natural way of finding out the evidence of possible variation
of this constant is by analyzing the atomic and molecular absorption spectra coming out of distant 
astronomical objects such as the high-red-shifted quasars \cite{Drinkwater, Webb1, Cowie, Bahcall}. 
The time scale at which these events are occurred corresponds to again 
$10^7 - 10^9$ years back and the statistical uncertainties in these systems can be reduced using 
the many-multiplet methods \cite{Murphy1, Sergei}. The impressive part of considering astrophysical
investigation of variation of $\alpha_e$ is that one can consider a large number of spectral lines for the 
analysis. As a matter of fact, the spectral lines from the highly charged ions can be investigated 
in this case which can have extra ordinary large enhancement of the relativistic effects \cite{berengut1,
berengut2}. In this paper, we analyze the relativistic sensitivity coefficients in Mn IX, Fe X, Co XI and 
Ni XII ions which are not explored before.

The considered Mn IX, Fe X, Co XI and Ni XII ions are of particular interest for analyzing their spectra 
as the ground states of these ions have the fine structure splittings. The transitions among these 
states can occur through the forbidden transitions and their wavelengths lie in the ultra-violet (UV)
region \cite{Nist}. The next excited levels are the $s$-states which can decay to the above two lower
states through the allowed channel. For an advantage, these transitions have wavelengths in the optical 
region \cite{Nist}. It can be noticed that the fine structure transitions have one more leading order
relativistic correction in $\alpha_e^2$ than the optical transitions. As a fact, the ratios of transition
frequencies in the above ions seem to be very promising quantities for carrying out the investigation of any
temporal variation of $\alpha_e$ by comparing these values from the spectra coming out of any astronomical objects 
with their corresponding laboratory values. It is identified from the solar extreme ultraviolet (EUV)
spectra line that most of them are the emission lines of Fe X ion \cite{Edlen, Dere, Thomas, Nussbaumer, 
Bhatia, Del Zanna}. These lines can be used to extract data for the electron density in the solar corona 
as proposed by Jordan \cite{Jordan}. Mason and Nussbaumer had also observed that under typical solar coronal 
condition, Cl like ions such as Fe X give rise to most of the prominent spectra \cite{Mason}. The red iron 
line corresponds to the  forbidden transition between the fine structure levels of the ground state of Fe X. 
The other ions Co XI and Ni XII are also important for the astrophysical study, but they are relatively less abundant 
in the astrophysical objects. Quite a few emission lines of Co XI are observed in the solar plasma, and in the spectra 
from theta-pinch plasma \cite{Thomas, Fawcett1, Fawcett2, Fawcett3}. Some of the lines of Co XI ranging the wavelengths
in between 65-340$\AA$ have been observed and tabulated in \cite{Peter}. It is also revealed from the 
data analysis of high-resolution soft X-ray spectrum of nearby F-type star Procyon that the emission 
lines contains spectra from Mn IX and Ni XII along with from other highly charged ions \cite{Raassen}. 
Moreover, these ions can be reproduced using the accelerators for their laboratory studies. Thus, theoretical
calculations of the spectral properties in these ions are necessary.

Theoretical determination of the atomic states in the considered ions are very challenging owing to the fact
that they contain five valence electrons in their outer most orbitals. There have been only few studies 
carried out for the evaluation of the lifetimes of the first excited states using semi-empirical, mean-field
and configuration interaction (CI) methods; however there are neither any theoretical calculations nor any 
observations of the lifetimes of the second excited states are available in the considered ions. In fact, 
the higher order relativistic effects are never investigated in these ions. We have developed here an all order 
perturbative method in the relativistic coupled-cluster (RCC) framework to carry out the study of 
correlation effects and relativistic corrections systematically in the first three low-lying states of the 
undertaken ions and would like to calculate precisely the electron detachment energies, the sensitivity coefficients for the 
variation of $\alpha_e$ involving the first three transitions and the lifetimes of the first two excited states. 
We shall also demonstrate the roles of the higher order relativistic effects in the estimation of the sensitivity 
coefficients.

\section{Theory and Method of Calculations}

\subsection{$\alpha_e$ Sensitivity Coefficient}

The energy expression for a state of any multi-electron atomic system can be approximated to \cite{Greiner}
\begin{eqnarray}
E_{n}\simeq c^2(Z\alpha_e)^2 \left \{\frac{1}{2n^2}+\frac{(Z\alpha_e)^2}{2n^3} \left (\frac{1}{|\kappa|}-\frac{3}{4n} \right ) \right \}  
\end{eqnarray}
where $Z$ is the atomic number of the atom, $c$ is the velocity of light, $n$ is the principal quantum number of the state and $\kappa = \pm(j+\frac{1}{2})$ is the relativistic 
quantum number with angular momentum of the state $j$. Since the relativistic effects to the energy
levels close to the nucleus are large due to the high angular velocity of the electron, the 
relativistic corrections to the energy levels can be approximated to \cite{Dzuba}
\begin{eqnarray}
\Delta=-\frac{Z^2_a}{2}\frac{(Z\alpha_e)^2}{\nu^3} \left ( \frac{1}{j+1/2}-\frac{Z_a}{Z\nu} \left [ 1-\frac{Z_a}{4Z} \right ] \right ), 
\end{eqnarray}
with $\nu$ is the effective principal quantum number and $Z_a$ is the effective charge
seen by an electron after the screening effect of the inner core electrons. As the atomic energy levels 
scale of the order of $\alpha_e^2$ after taking the relativistic correlation effects into account, hence 
the transition frequencies among the atomic levels are very sensitive to a small change in $\alpha_e$ value
which will, obviously, get enhanced for a large atomic number $Z$ and for a small value of $\nu$. Therefore, this 
sensitivity is large in the highly charged ions. For the theoretical investigation, this sensitivity can be
estimated by considering a relativistic method to calculate transition frequency ($\omega$) of a transition by
expressing them into
\begin{eqnarray}
\omega(x)\approx\omega_0+qx 
\end{eqnarray}
where $\omega_0$ corresponds to transition frequency with the laboratory value of the fine structure constant 
$\alpha_0$,   
$x=(\frac{\alpha_e}{\alpha_0})^2-1$ is the Taylor coefficient of the first derivative of $\omega$ and $q=\frac{d \omega}{d x}|_{x=0}$ 
is known as the sensitivity coefficient for the variation of the fine structure constant. For the numerical estimate of
the $q$-factor, it can be evaluated at the first order correction in $\alpha_e^2$ by
\begin{eqnarray}
q\approx\frac{\omega(+ x)-\omega(- x)}{2 x},
\end{eqnarray}
for a given choice of small value of $x$ which, in the present calculation, is chosen as $0.05$.
 
 \subsection{Lifetime of an Atomic State}

The transition probabilities due to the E1, M1 and E2 channels of an atomic transition $| \Psi_f \rangle \rightarrow 
| \Psi_i \rangle$ are given by \cite{Johnson}
\begin{eqnarray}
&& A^{E1}_{fi} =  \frac{2.0261\times 10^{-6}}{\lambda_{fi}^3 g_f} S_{fi}^{E1}, \label{eqn5} \\
&& A^{E2}_{fi} = \frac{1.1195\times 10^{-22}}{\lambda_{fi}^5 g_f} S_{fi}^{E2} \label{eqn6} \\
\text{and}  && \nonumber \\
&& A^{M1}_{fi} =  \frac{2.6971\times 10^{-11}}{\lambda_{fi}^3 g_f} S_{fi}^{M1}, \label{eqn7} \ \ \ \ \ \ \
\end{eqnarray}
where the quantity $S^O_{fi} = \mid {\langle \Psi_f \vert \vert O \vert \vert \Psi_i \rangle} \mid^2$ 
is known as the line strength of a transition for a corresponding channel $O$ and is estimated here 
in atomic unit (au), the transition wavelength $\lambda_{fi}$ is taken in $cm$ and $g_f=2J_f+1$
is the degeneracy factor with the angular momentum $J_f$ of the state $| \Psi_f \rangle$. The determined
transition probabilities are obtained in $s^{-1}$ from the above quantities.

The emission (absorption) oscillator strengths $f_{fi}$ ($f_{if}$) due to the above transition 
probabilities are given by \cite{Sobelman}
\begin{eqnarray}
f_{fi} = 1.4992\times 10^{-24} A_{fi} \frac{g_f}{g_i}\lambda_{fi}^2
\label{eqn3}
\end{eqnarray}
which follows that $g_i f_{if} = - g_f f_{fi}$.

The reduced matrix elements for the single particle orbitals corresponding to E1,
M1 and E2 transitions are given by \cite{Johnson}
\begin{eqnarray}
\langle \kappa_f\, ||\,e1\,||\, \kappa_i \rangle &=&
\langle \kappa_f\, ||\,C^{(1)}\,||\,\kappa_i \rangle \nonumber \\ 
&& \int_0^{\infty} dr \ r \ (P_f(r)P_i(r)+Q_f(r)Q_i(r)), \ \ \ \ \ \ \
\end{eqnarray}

 \begin{center}
\begin{figure}[t]
\includegraphics[width=6.5cm,height=4.0cm,clip=true]{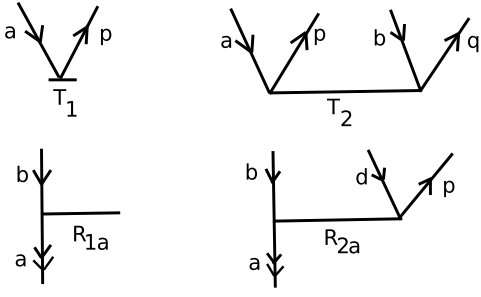}
\caption{Goldstone diagrammatic representation of the $T$ and $R_a$ operators.
Horizontal lines represent the reference state $|\Phi_0 \rangle$, line with index $p$ going up from the reference means creation
of an electron in the virtual $p$ orbital (similarly line $p$ coming into the reference line from below means electron annihilation 
from a virtual orbital $p$ as used in the latter figures), line with index $a$ going down from the reference means annihilation of the electron
from the occupied orbital $a$ and $b$ line coming to the reference means it is creating an electron in the occupied orbital $b$.}
\label{fig1}
\end{figure}
\end{center}

  \begin{eqnarray}
\langle \kappa_f\, ||\,m1\,||\, \kappa_i \rangle &=&  \frac {(\kappa_f+\kappa_i)}
{\alpha}\langle - \kappa_f\, ||\,C^{(1)}\,||\,\kappa_i \rangle \nonumber \\ 
&& \int_0^{\infty} dr \ r \ (P_f(r)Q_i(r)+Q_f(r)P_i(r)), \ \ \ \ \ \ \
\end{eqnarray}
and 
\begin{eqnarray}
\langle \kappa_f\, ||\,e2\,||\,\kappa_i \rangle &=& \langle \kappa_f\, ||\,C^{(2)}\,||\, \kappa_i \rangle \nonumber \\
&& \int_0^{\infty} dr \ r^2 \ (P_f(r)P_i(r)+Q_f(r)Q_i(r)), \nonumber  \\
\end{eqnarray}
where $P(r)$ and $Q(r)$ denote the large and small components of the radial parts of the single
particle Dirac orbitals, respectively. The reduced Racah coefficients are given by
\begin{eqnarray}
\langle \kappa_f\, ||\, C^{(k)}\,||\, \kappa_i \rangle &=& (-1)^{j_f+1/2} \sqrt{(2j_f+1)(2j_i+1)} \ \ \ \ \ \ \ \ \nonumber \\
                  &&        \left ( \begin{matrix}
                              j_f & k & j_i \cr
                              1/2 & 0 & -1/2 \cr
                                       \end{matrix}
                            \right ) \pi(l_{\kappa_f},k,l_{\kappa_i}), \ \ \ \ \
\end{eqnarray}
with
\begin{eqnarray}
\pi(l,k,l') &=&
\left\{\begin{array}{ll}
\displaystyle
1 & \mbox{for } l+k+l'= \mbox{even}
\\ [2ex]
\displaystyle
0 & \mbox{otherwise.}
\end{array}\right.
\label{eqn12}
\end{eqnarray}

\begin{center}
\begin{figure}[t]
\includegraphics[width=7.5cm, height=9.5cm, clip=true]{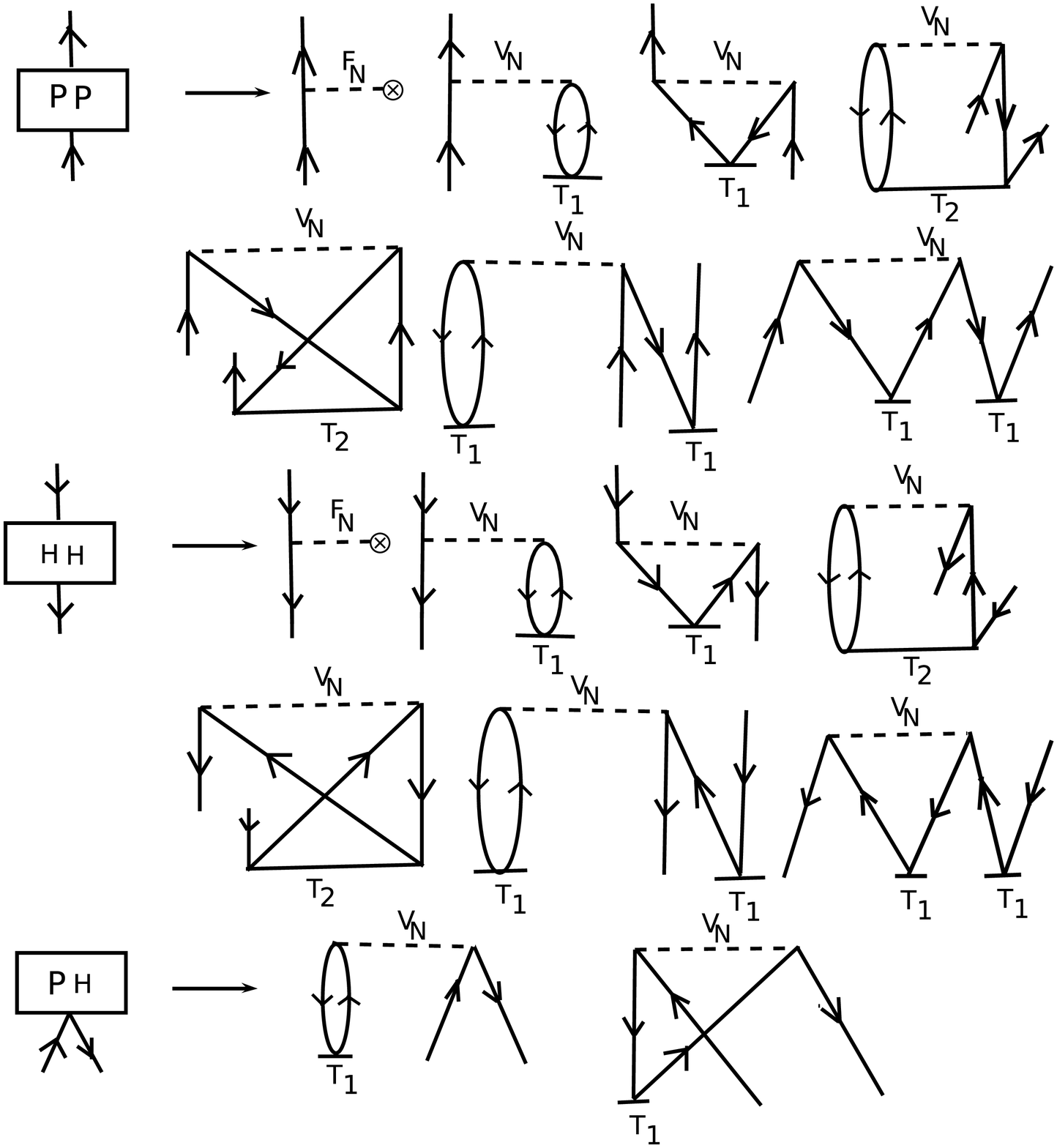}
\caption{Effective one-body diagrams constructed from $\widehat{H_Ne^T}$ for the evaluation of the electron detachment energy and the $R_a$ amplitudes.} 
\label{fig2}
\end{figure}
\end{center}

The lifetime of a given atomic state is the inverse of the total transition 
probabilities involving all possible spontaneous emission channels. i.e. 
the lifetime (in $s$ corresponding to the aforementioned units) of the state 
$|\Psi_k \rangle$ is given by
\begin{eqnarray}
\tau_k &=& \frac {1} {\sum_{O,i} A^{O}_{ki}},
\label{eqn8}
\end{eqnarray}
where the sum over $O$ represents all possible decay channels due to the transition 
operators $O$ and the sum over $i$ corresponds to all the lower transition states. 

\begin{center}
\begin{figure}[t]
\includegraphics[width=8.3cm, height=8.5cm, clip=true]{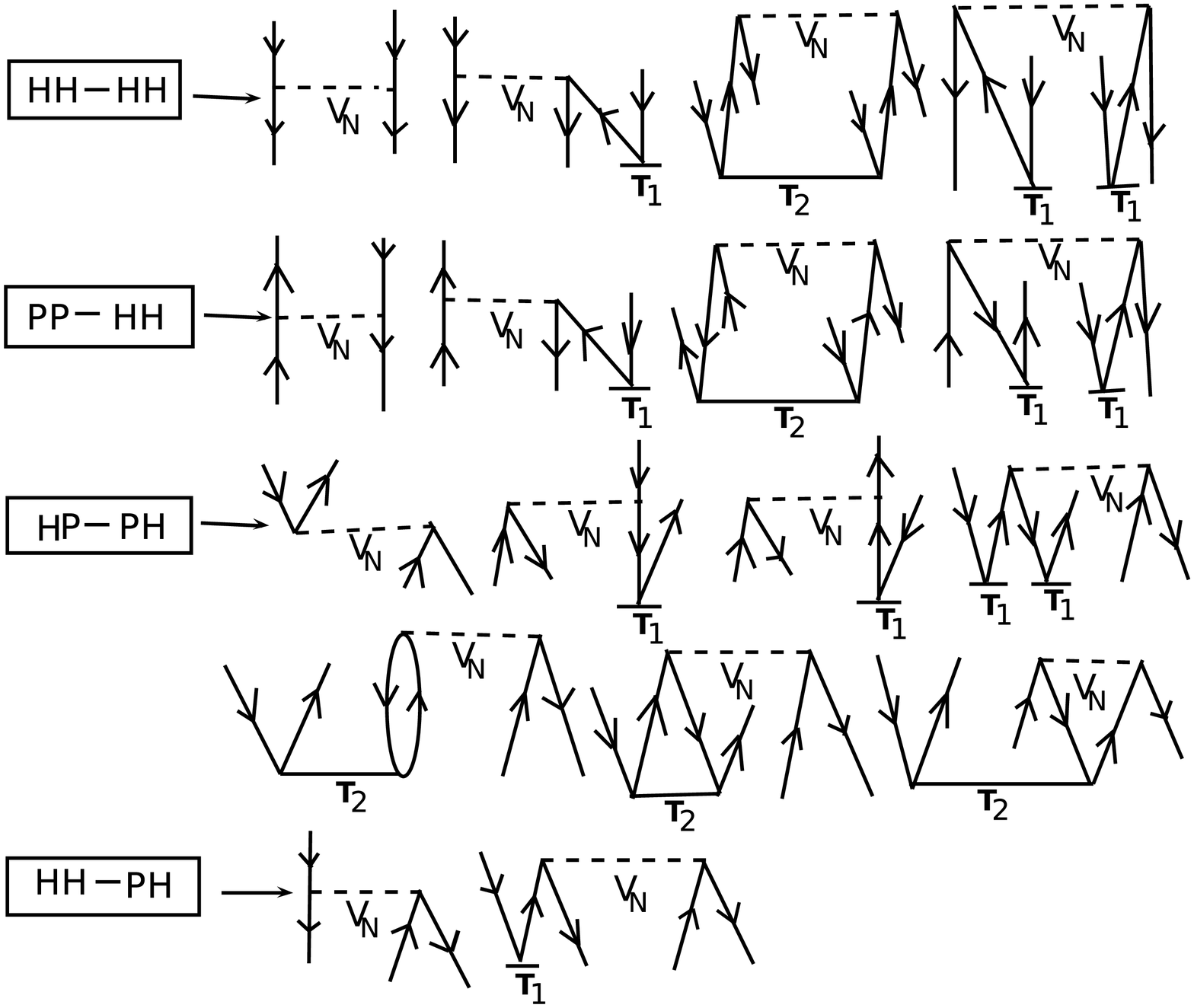}
\includegraphics[width=8.5cm, height=8.5cm, clip=true]{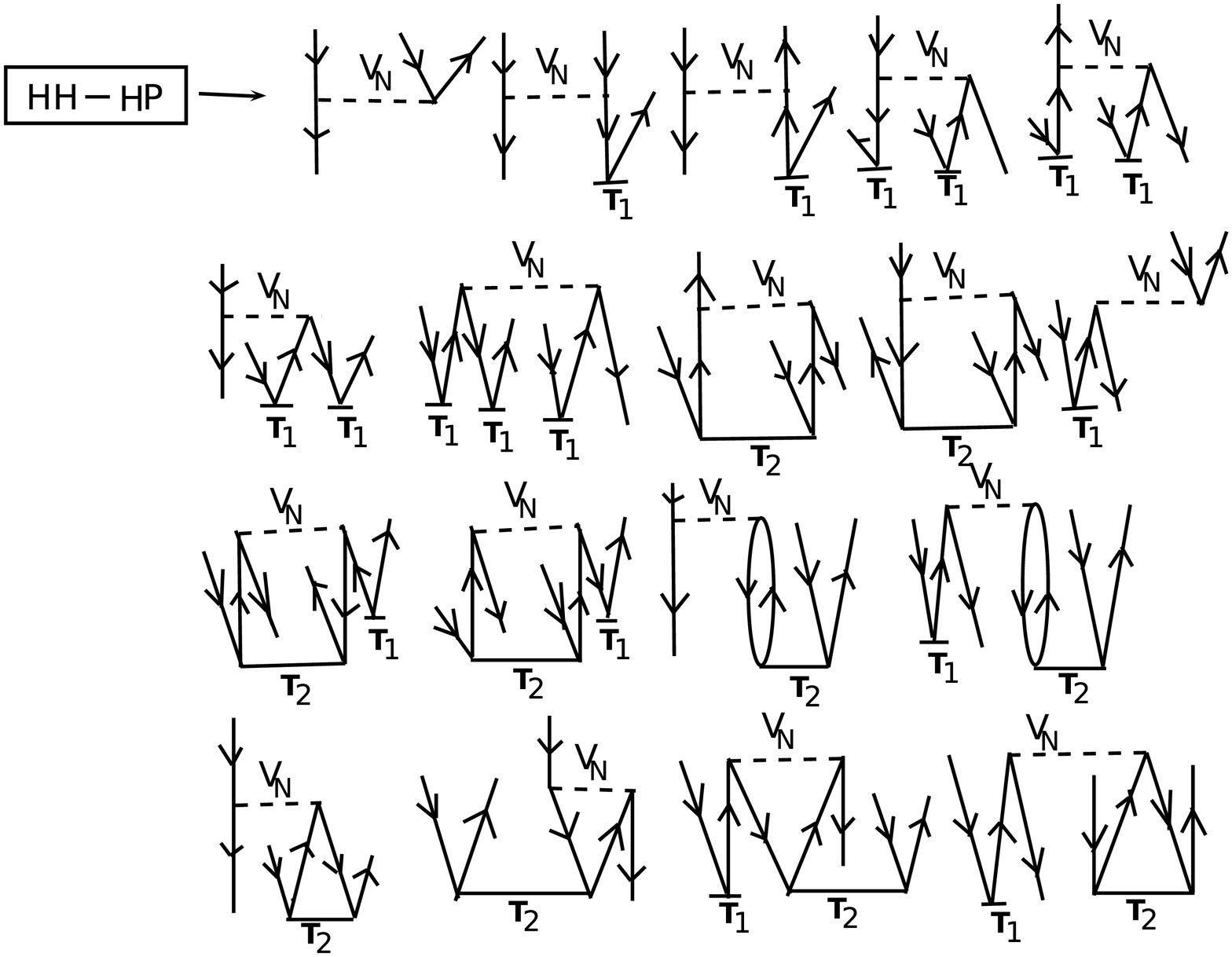}
\caption{Effective two-body diagrams constructed from $\widehat{H_N e^T}$ for the calculation of the electron detachment energy and the $R_a$ amplitudes.} 
\label{fig3}
\end{figure}
\end{center}

\subsection{RCC method for electron detachment}

The ground and the first two excited states, those are of particular interest in the present work, 
of the considered ions have a structure of one electron less than the closed-shell configuration
$[3s^2 \ 3p^6]$. These states can be generated 
by removing one electron from the respective $3p_{3/2}$, $3p_{1/2}$ and $3s$ orbitals in three
separate steps. Construction of the atomic state function (ASF) allowing couplings between all possible 
configuration state functions (CSFs) having the same angular momentum is not easily viable in these ions 
owing to the presence of many electrons in the valence space. One of the approachable  
ways of calculating these ASFs is to evaluate wave function for the $[3s^2 \ 3p^6]$ configuration 
by accounting correlations among all these electrons and later remove an electron from the respective orbital
in a Fock-space representation. In this procedure, one has the flexibility to use the reduced matrix 
elements for minimizing the computational requirements so that it can afford to include the correlation effects
more efficiently.

\begin{center}
\begin{figure}[t]
\includegraphics[width=7.5cm, height=4.5cm, clip=true]{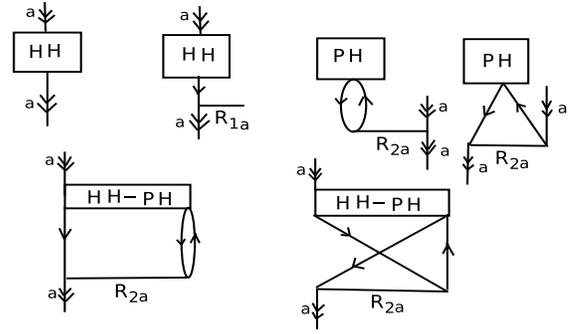}
\caption{Diagrams contributing to the calculations of the detachment energies. The line with double arrow means the orbital $a$ from which an electron has been removed.} 
\label{fig4}
\end{figure}
\begin{figure}[t]
\includegraphics[width=7.5cm, height=4.5cm, clip=true]{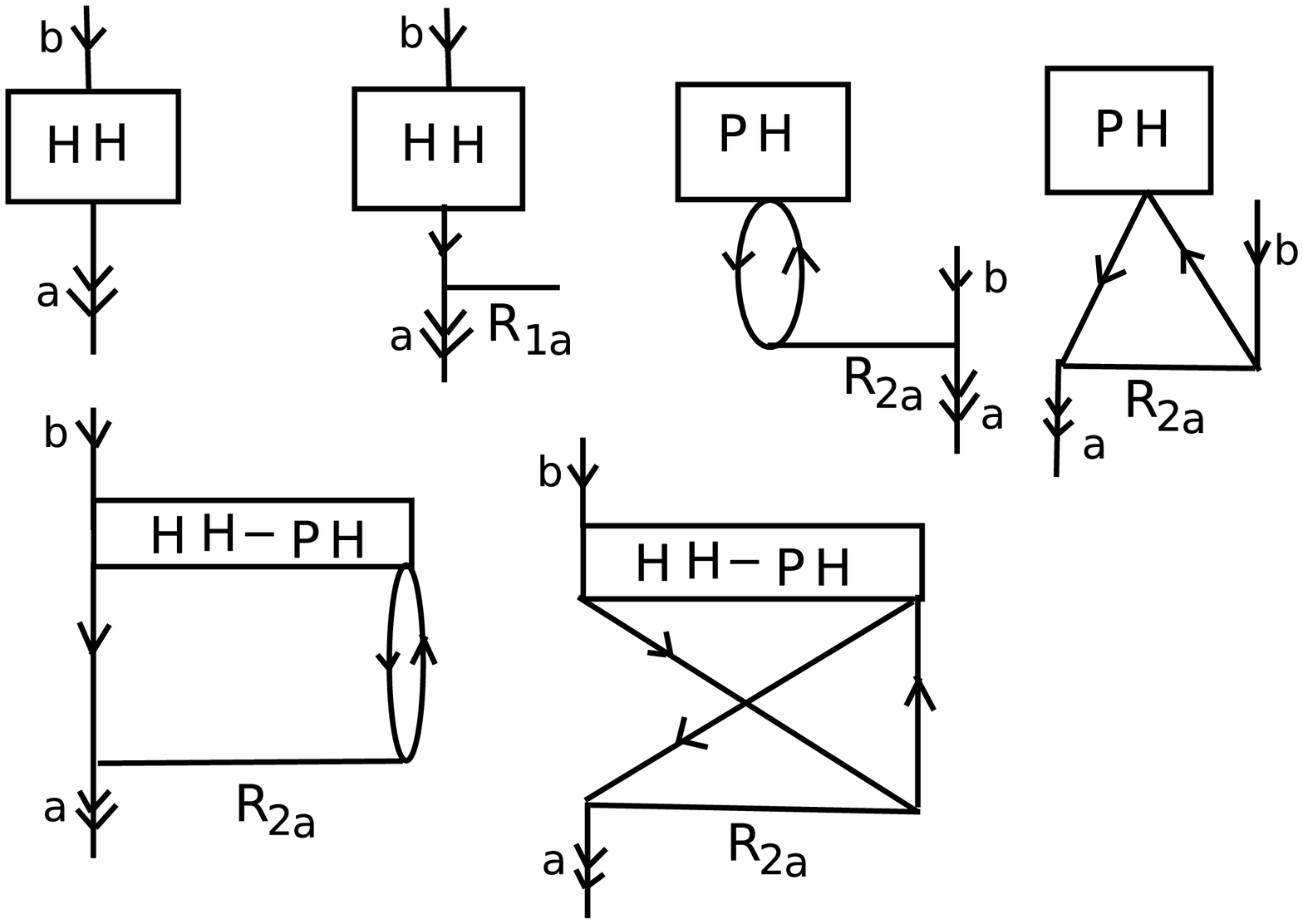}
\caption{Diagrams for the evaluation of the $R_{1a}$ amplitudes. The first diagram arises from the right hand side (rhs)
and the rest are from the left hand side (lhs) of Eq. (\ref{eq26}).} 
\label{fig5}
\end{figure}
\end{center}

\begin{center}
\begin{figure}[t]
\includegraphics[width=7.5cm, height=7.5cm, clip=true]{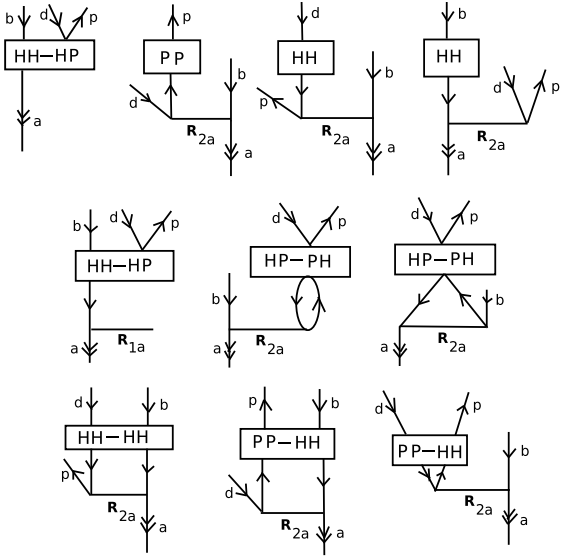}
\caption{Diagrams for the evaluation of the $R_{2a}$ amplitudes. The first diagram arises from the right hand side (rhs)
and the rest are from the left hand side (lhs) of Eq. (\ref{eq27}).}     
\label{fig6}
\end{figure}
\end{center}
The other important aspect is to construct a perturbative method for the inclusion of the correlation 
effects among the electrons to infinite order for the high precision calculations of the 
wave functions. Keeping in mind the above mentioned complexity in the mixing of CSFs in order
to obtain the desired ASFs in the considered ions, the developed ($Z$-1) Fock-space based coupled-cluster
method in this work in the relativistic frame work (here onwards we refer it to RCC method) seems to be 
one of the most elegant approaches to employ in a closed-core with $Z$ number of electrons and generate
the atomic states of its ion by removing one electron. To describe this procedure, we express the above 
ASFs in the form
\begin{eqnarray}
|\Psi_a \rangle &=& a_a |\Psi_0 \rangle + R_a a_a |\Psi_0 \rangle ,
\end{eqnarray}
where $|\Psi_0 \rangle$ represents ASF of the $[3s^2 \ 3p^6]$ configuration and $a_a$ represents annihilation 
of the electron from the $a$ orbital of the closed-core. The important
point to be noticed here is that we have already accounted correlations among all the electrons from 
the $[3s^2 \ 3p^6]$ configuration in the construction of $|\Psi_0 \rangle$.
The $R_a$ operator, thus, need to remove the extra correlation effects that is being 
taken into account for the extra $a_a$ electron in the determination of $|\Psi_a \rangle$. In the 
RCC ansatz, the above expression yields to
\begin{eqnarray}
|\Psi_a \rangle &=& a_a e^T |\Phi_0 \rangle + R_a a_a e^T |\Phi_0 \rangle  \nonumber \\
                &=& e^T (1+ R_a) a_a |\Phi_0 \rangle  \nonumber \\
                &=& e^T (1+ R_a) |\Phi_a \rangle,
\end{eqnarray}
where $|\Phi_0 \rangle$ is a mean-field wave function for the closed-core which we obtain by using
the Dirac-Hartree-Fock (DHF) method, $T$ is the RCC operator that 
accounts correlation effects in terms of generating all possible CSFs from $|\Phi_0 \rangle$
and $|\Phi_a \rangle = a_a |\Phi_0 \rangle$ is defined as the modified reference state for the 
new ASF $|\Psi_a \rangle$.

The considered ions are highly charged systems, but the electron correlation effects in these ions are anticipated 
to dominate over the quantum electrodynamics (QED) interactions. However, these QED effects will be
immensely large compared to the neutral atoms and the singly charged ions. For the highly charged
ions, the orbitals are contracted and concentrated around the nucleus. In such case, the many-body
atomic Hamiltonian can be approximated to the kinetic energies of the electrons expressed using the 
Dirac theory, the nuclear potential and the leading order correction terms from QED. This 
would be a quite reasonable choice for describing the relativistic effects in the bound electrons
of the considered ions. In this work, we restrict the two-body interactions between the electrons
to one-photon exchange interaction due to the longitudinal and transverse components as encapsulated 
in terms of the Coulomb and the approximated frequency independent Breit interactions, respectively.
All together, the atomic Hamiltonian is given by
\begin{eqnarray}
H&=& \sum_i \Lambda_i^+ \left [ c\mbox{\boldmath$\alpha$}_i\cdot \textbf{p}_i+(\beta_i -1)c^2 + V_{n}(r_i) \right ] \Lambda_i^+ \nonumber \\ && +
\sum_{i,j>i} \Lambda_i^+ \Lambda_j^+ V_{ee}(r_{ij}) \Lambda_i^+ \Lambda_j^+
\end{eqnarray}
where $\mbox{\boldmath$\alpha$}_i$ and $\beta_i$ are the usual Dirac matrices
and the symbol $\Lambda^+$ ensures that when the operators act only on the positive 
energy states, it gives the finite values else the contributions from the negative energy
states are suppressed. Subtraction of the identity operator from $\beta$ means that the energies are scaled 
over the rest mass energies of the electrons. We take the effective nuclear potential as
$V_n(r_i)=V_{fm} (r_i) + V^{VP}(r_i) + V^{SE}(r_i)$ and the two-body interaction potential
as $V_{ee}(r_{ij})= \frac{1}{r_{ij}} + V_B(r_{ij})$.
Therefore, in the Dirac-Coulomb (DC) approximation we have 
\begin{eqnarray}
H^{DC}&=& \sum_i \left [ c\mbox{\boldmath$\alpha$}_i\cdot \textbf{p}_i+(\beta_i -1)c^2 + V_{fm}(r_i) +
\sum_{j>i} \frac{1}{r_{ij}} \right ], \nonumber \\
\end{eqnarray}
with $V_{fm}(r_i)$ is the nuclear Coulomb potential obtained using the Fermi charge distribution.
The approximated frequency independent Breit interaction Hamiltonian is given by \cite{Breit}
\begin{eqnarray}
V_B(r_{ij})=-\frac{1}{2r_{ij}}\{\mbox{\boldmath$\alpha$}_i\cdot \mbox{\boldmath$\alpha$}_j+
(\mbox{\boldmath$\alpha$}_i\cdot\bf{\hat{r}_{ij}})(\mbox{\boldmath$\alpha$}_j\cdot\bf{\hat{r}_{ij}}) \}. 
\end{eqnarray}
The leading order corrections from the vacuum polarization (VP) radiative effects is taken 
to be the Uehling and Wichmann-Kroll potential as $V^{VP}(r)=V_{Uhl}(r)+V_{WK}(r)$ \cite{Flambaum} with
\begin{eqnarray}
V_{Uhl}(r)&=&  - \frac{4}{9 c \pi} V_{fm}(r)
\int_1^{\infty}dt\sqrt{t^2-1}\left(\frac{1}{t^2}+\frac{1}{2t^4}\right) e^{-2ctr} \nonumber \\
\end{eqnarray}
and
\begin{eqnarray}
V_{WK}(r)&=&-\frac{2}{3}\frac{1}{c\pi}V_{fm}(r) \frac{0.092 c^2 Z^2}{1+(1.62 cr)^4}.
\end{eqnarray}
Similarly, the self-energy (SE) correction from the radiative effect is approximated
to the contributions from the magnetic and electronic form factors as $V^{SE}(r)=V_{mf}(r)+V_{ef}(r)$ \cite{Flambaum} with
\begin{eqnarray}
V_{mf}(r)=-\frac{1}{4 c^2 \pi } \mbox{\boldmath$\gamma$}.\mbox{\boldmath$\nabla$} 
\left [V_{fm}(r)\left (\int^{\infty}_{1}dt \frac{e^{-2ctr}}{\sqrt{t^2-1}}\right)\right] 
\end{eqnarray}
and
\begin{eqnarray}
V_{ef}(r)&=&-A(Z,r)\frac{1}{c\pi}V_{fm}(r) \int^{\infty}_1 dt \frac{e^{-2ctr}}{\sqrt{t^2-1}}[\left( 1-\frac{1}{2t^2}\right)\nonumber \\
&&\times \{ln(t^2-1)+4 ln(c/Z+0.5)\}-\frac{3}{2}+\frac{1}{t^2}]\nonumber \\
&& - B(Z) \frac{Z^4}{c^3} e^{-Zr} ,  
\end{eqnarray}
for the quantities $A(Z,r)=[1.071-1.97((Z-80)/c)^2-2.128((Z-80)/c)^3+0.169((Z-80)/c)^4]cr/(cr+0.07Z^2/c^2)$ and $B(Z)=0.074+0.35Z/c$.

In the above expressions, we have adopted au units which we shall follow-up in the rest of the paper.
Also it is assumed that the mass of the nucleus is infinitely heavy. Thus, the corrections from the 
reduced mass of the electrons and the nuclear recoil effect, which are inversely proportional to the
nuclear mass \cite{Johnson,Shabaev}, are neglected in the present calculations.

\begin{center}
\begin{figure}[t]
\includegraphics[width=8.5cm, height=4.5cm, clip=true]{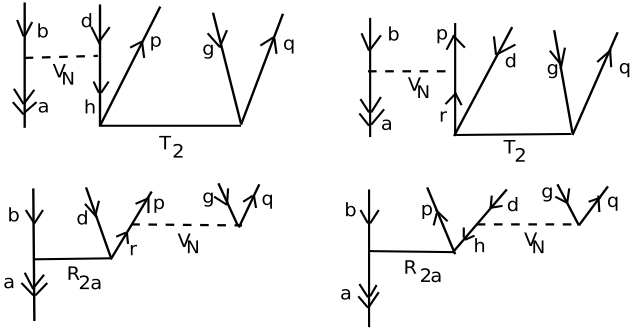}
\caption{Goldstone diagrams of the $R_{3a}^{pert}$ operator that are accounted perturbatively in the CCSD(T) method.} 
\label{fig7}
\end{figure}
\end{center}
For the simplicity, the normal order Hamiltonian has been used in our calculations with respect to the reference state 
$|\Phi_0 \rangle$ by defining
\begin{eqnarray}
H_N=H-\langle\Phi_0|H|\Phi_0\rangle=H - E_{SCF},
\end{eqnarray}
with the self-consistent-field (SCF) Hartree-Fock energy $E_{SCF}$.
 
The amplitude solving equations for the $T$ operators for a closed-shell configuration are well known 
and can be referred to \cite{Bartlett, Bijaya, Mukherjee} for any more required explanation. We have restricted to only
singly and doubly excited configurations from $|\Phi_0 \rangle$ in our calculations (known as CCSD method)
by defining $T=T_1 + T_2$ which in the second quantization notation are given by
\begin{eqnarray}
   T_1 =\sum_{a,p}a^{\dagger}_p a_a t^p_a, 
\ \ \ \text{and} \ \ \ T_2 =\frac{1}{4}\sum_{ab,pq}a^{\dagger}_pa^{\dagger}_qa_ba_a t^{pq}_{ab},
\end{eqnarray}
where the subscripts $a,b$ and $p,q$ represent the core and virtual orbitals, $a$ and $a^{\dagger}$ are the annihilation 
and creation operators, and $t_a^p$ and $t_{ab}^{pq}$ are the excitation amplitudes for the creation of the singly 
and doubly excited configurations. 

Now the eigenvalue equations for the required states are given by
\begin{eqnarray}
H |\Psi_a \rangle &=& E_a |\Phi_a \rangle \nonumber \\ 
( (\widehat{H_N e^T})_{fc} + (\widehat{H_N e^T})_{op} &+& E_{SCF} )  \{1+R_a\}|\Phi_a \rangle \nonumber \\ &=& E_a \{1+R_a\}|\Phi_a \rangle  \nonumber \\ 
(\widehat{H_N e^T})_{op} \{1+R_a\}|\Phi_a \rangle &=& \Delta E_a \{1+R_a\}|\Phi_a \rangle.
\end{eqnarray}
The subscripts $fc$ and $op$ represent the fully contracted and operator form of $\widehat{H_N e^T}$, with widehat symbol
representing only the connecting terms, that are obtained multiplying by $e^{-T}$ from the left hand side in the above 
equation and
$\Delta E_a$ refers to the electron detachment energy or ionization potential (IP) of the electron to remove it
from the orbital $a$ of the $|\Psi_0 \rangle$ state; i.e. from the ASF of the $[3s^2 \ 3p^6]$ configuration. Here onwards
we drop the subscript $op$ for further discussions as the fully contracted terms will not appear any more.

\begin{center}
\begin{figure}[t]
\includegraphics[width=8.5cm,clip=true]{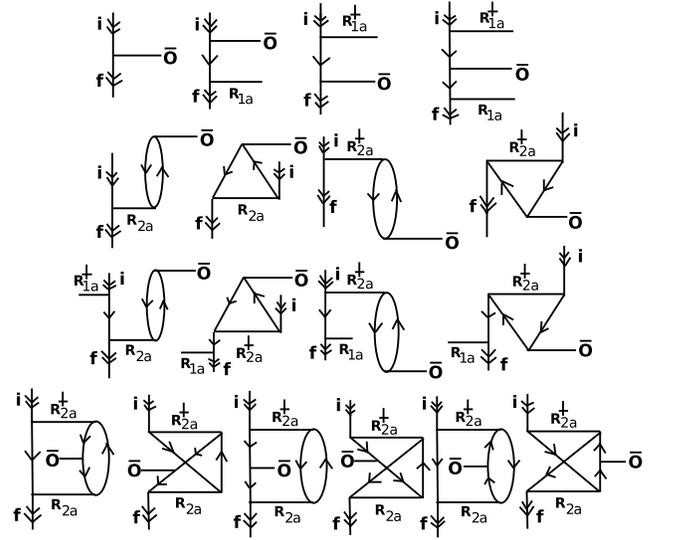}
\caption{Final property evaluating diagrams connecting effective one-body $\overline{O}$ operators with $R_a$ and its complex conjugate ($cc$) diagrams.}  
\label{fig8}
\end{figure}
\end{center}

\begin{center}
\begin{figure}[t]
\includegraphics[width=8.5cm,clip=true]{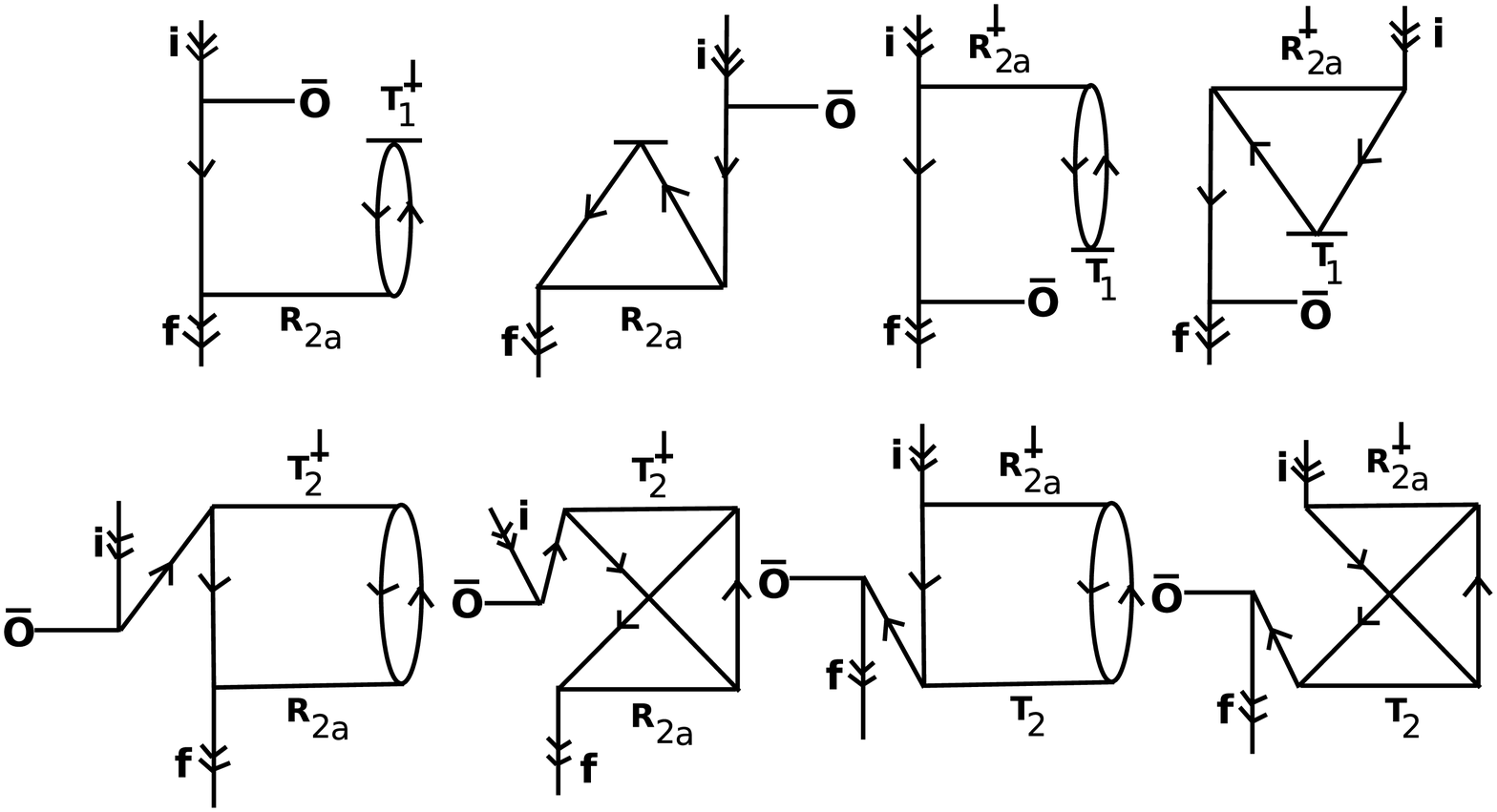}
\caption{Few important diagrams connecting effective two-body $\overline{O}$ operators with $R_a$ and its $cc$ diagrams that 
are contributing significantly in the transition amplitude calculations.}
\label{fig9}
\end{figure}
\end{center} 

\begin{table*}[t]
\caption{Electron detachment energies (in $cm^{-1}$) of few low lying states in Mn IX, Fe X, Co XI and Ni XII ions using
the DHF, CCSD and CCSD(T) methods with DC, (DC+Breit), (DC+Breit+VP) and (DC+Breit+VP+SE) Hamiltonians. The results are compared with the experimental values 
listed in the NIST database \cite{Nist}. The differences between the NIST data and our final results from the CCSD(T) 
method are quoted as $\delta$ in percentage (\%). We also give $y=(\Delta E_a^{DC} - \Delta E_a^{final})/\Delta E_a^{final}$ values
for the undertaken states in the above ions to demonstrate the trends of the relativistic effects in the heavier ions.}
\begin{ruledtabular}
\begin{tabular}{lccccccccc}
Method& \multicolumn{1}{c}{$3s^23p^5 \ ^2P_{3/2} $} & \multicolumn{1}{c}{$3s^23p^5 \ ^2P_{1/2} $}&\multicolumn{1}{c}{$3s3p^6   \ ^2S_{1/2} $} \\
           &  CCSD(T)               &  CCSD(T)              &  CCSD(T)   \\
\hline \\
 Mn IX & & \\

DHF        &1575067.58              &1588266.78              &1909515.85    \\
DC         &1576861.85 ($-471.09$)  &1589743.76 ($-490.04$)  &1843896.78 ($-1167.62$)  \\
$+$Breit   &1576457.42 ($-475.16$)  &1589006.95 ($-496.11$)  &1843351.89 ($-1169.82$)   \\
$+$VP      &1576453.12 ($-475.17$)  &1589002.80 ($-496.11$)  &1843377.15 ($-1169.76$)  \\
$+$SE      &1577268.24 ($-472.06$)  &1589853.50 ($-492.94$)  &1844714.05 ($-1163.77$)   \\
& & \\
NIST        &1576600.01             &1589146.02              &1842008.01              \\
$\delta$(\%)&0.04                   &0.04                    &0.14                   \\
$y$         &0.0002                 &0.0001                  &0.0004                 \\
\hline \\
 Fe X & & \\
                    
DHF        &1883771.66              &1900218.87              &2245974.19                 \\
DC         &1886047.50 ($-512.23 $) &1902131.24 ($-534.99 $) & 2176829.41 ($-1309.54 $)     \\
$+$Breit   &1885528.83 ($-518.04 $) &1901215.38 ($-542.47 $) & 2176152.87 ($-1312.49 $)     \\
$+$VP      &1885523.73 ($-518.06 $) &1901210.48 ($-542.48 $) &2176184.13  ($-1312.43 $)    \\
$+$SE      &1886621.73 ($-514.53 $) &1902361.90 ($-538.86 $) &2177937.43  ($-1305.17 $)     \\
& & \\
NIST        &1884000.01             &1899683.01              &2173249.02                \\
$\delta$(\%)&0.14                   &0.14                    &0.21                      \\
$y$         &0.0003                 &0.0001                  &0.0005                    \\
\hline \\
Co XI & & \\

DHF        &2217435.11              &2237682.21              &2608011.58                     \\
DC         &2219848.89 ($-509.21 $) &2239675.97 ($-535.29 $) &2534877.86 ($-1457.54 $)        \\
$+$Breit   &2219196.01 ($-515.14 $) &2238552.11 ($-542.92 $) &2534047.65 ($-1460.75 $)       \\
$+$VP      &2219190.05 ($-515.15 $) &2238546.42 ($-542.94 $) &2534085.85 ($-1460.71 $)        \\
$+$SE      &2220636.28 ($-511.12 $) &2240070.27 ($-538.77 $) &2536349.04 ($-1451.66 $)        \\
& &  \\
NIST        &2221000.01             &2240345.02              &2534630.02                  \\
$\delta$(\%)&0.02                   &0.01                    &0.06                        \\
$y$         &0.0003                 &0.0002                  &0.0006                        \\

\hline \\
 Ni XII & & \\

DHF        &2575996.48             &2600654.35              &2995633.72                     \\
DC         &2579954.22 ($-467.87 $)&2604148.24 ($-496.73 $) &2919968.40 ($-1496.84 $)        \\
$+$Breit   &2578738.04 ($-605.35 $)&2602370.96 ($-637.96 $) &2918845.93 ($-1485.21 $)       \\
$+$VP      &2578731.71 ($-604.76 $)&2602365.04 ($-637.34 $) &2918893.03 ($-1484.39 $)        \\
$+$SE      &2580601.25 ($-600.73 $)&2604345.11 ($-632.42 $) &2921767.82 ($-1474.02 $)        \\
&  & \\
NIST        &2577200.01            &2600829.02              &2915815.02                     \\
$\delta$(\%)&0.13                  &0.13                    &0.20                           \\
$y$         &0.0002                &0.0001                  &0.0006                          \\

\end{tabular}
\end{ruledtabular}
\label{tab1}
\end{table*}

We again restrict $R_a$ operators to account only the singly and doubly excited configurations from 
the corresponding $|\Phi_a \rangle$ reference states in order to be consistent with the CCSD method
by defining $R_a=R_{1a}+ R_{2a}$ which in the second quantization form are expressed as
\begin{eqnarray}
   R_{1a} =\sum_{b\ne a}a^{\dagger}_b a_a r^b_a, 
\ \ \text{and} \ \ R_{2a} =\frac{1}{2}\sum_{bd,p}a^{\dagger}_ba^{\dagger}_pa_d a_a r^{bp}_{ad}, \ \
\label{rad}
\end{eqnarray}
where the sum for $R_{2a}$ includes $b=a$ without any loss of generality to facilitate the simple angular momentum 
algebra at the cost of violating the Pauli's exclusion principle. The extra contributions anticipated from
these unphysical contributions are, however, cancel out from the direct and exchange parts of the
two-body diagrams by allowing contributions only from the linked diagrams in the calculations \cite{Lindgren}. Goldstone 
diagrammatical interpretation of the $T_1$, $T_2$, $R_{1a}$ and $R_{2a}$ operators are shown in Fig. \ref{fig1}.

The energy and amplitude solving equations for the $R_a$ wave operators are given by
\begin{eqnarray}
\langle \Phi_a| \widehat{H_N e^T} \{1+R_a\}|\Phi_a\rangle &=& \Delta E_a \\
\langle \Phi^b_a| (\widehat{H_N e^T}-\Delta E_a) R_a |\Phi_a \rangle &=&  - \langle \Phi^b_a| \widehat{H_N e^T} |\Phi_a \rangle \label{eq26}
\end{eqnarray}
and 
\begin{eqnarray}
\langle \Phi^{pb}_{da}| (\widehat{H_N e^T}-\Delta E_a) R_a |\Phi_a \rangle &=&  -\langle \Phi^{pb}_{da}|\widehat{H_N e^T} |\Phi_a \rangle, \label{eq27} \ \ \ \ \
\end{eqnarray}
where $|\Phi_a^b \rangle$ are the
singly excited configurations from $|\Phi_a \rangle$ constructed as replacing orbital $a$ by orbital $b$ and $|\Phi^{pb}_{da} \rangle$
denotes doubly excited configurations from $|\Phi_a \rangle$ constructed as replacing orbital $a$ by orbital $b$ along with exciting an electron
from the occupied orbital $d$ to virtual orbital $p$ as per the definitions given in Eq. (\ref{rad}). The above
non-linear equations are solved self-consistently along with its energy evaluating equation. 

\begin{table}[t]
\caption{Sensitivity coefficients $q$ (in $cm^{-1}$) for the first three low-lying transitions in Mn IX, Fe X, Co XI and Ni XII ions from 
the CCSD(T) method using the DC, (DC+Breit), (DC+Breit+VP) and (DC+Breit+VP+SE) Hamiltonians.}
\begin{ruledtabular}
\begin{tabular}{lcccc}
Method& \multicolumn{1}{c}{$3s^23p^5   \ ^2P_{3/2}\rightarrow $} & $3s^23p^5   \ ^2P_{1/2} \rightarrow$ & $3s^23p^5     \ ^2P_{3/2} \rightarrow$ \\
      & \multicolumn{1}{c}{$3s3p^6     \ ^2S_{1/2}$}             & $3s3p^6     \ ^2S_{1/2}$             & $3s^23p^5     \ ^2P_{1/2}$      \\

\hline \\
 Mn IX & & \\
DHF       &18387.57  &4934.39   &13444.18     \\
DC        &16120.40  &3014.10   &13105.50      \\
$+$Breit  &15933.81  &3160.00   &12772.97      \\
$+$VP     &15977.78  &3204.45   &12773.33       \\
$+$SE     &16849.90  &3984.50   &12865.40       \\

\hline \\
 Fe X & & \\
DHF       &22613.10   &5832.88  &16780.22      \\
DC        &19898.70   &3505.90  &16392.70      \\
$+$Breit  &19694.16   &3698.94  &15995.22      \\
$+$VP     &19748.35   &3752.61  &15995.74      \\
$+$SE     &20834.87   &4703.69  &16131.18      \\

\hline \\
 Co XI & & \\
DHF       &27516.40   &6823.61   &20692.78      \\
DC        &24273.10   &4026.20   &20246.90       \\
$+$Breit  &24050.83   &4276.94   &19773.89       \\
$+$VP     &24116.81   &4342.19   &19774.62       \\
$+$SE     &25458.00   &5488.80   &19969.20        \\

\hline \\
DHF       &33167.97  &7921.84    &25246.13      \\
 Ni XII & & \\
DC        &29338.30  &4594.00    &24746.20       \\
$+$Breit  &29101.08  &4909.73    &24191.34       \\
$+$VP     &29180.79  &4988.45    &24192.34       \\
$+$SE     &30828.80  &6363.90    &24464.90       \\

\end{tabular}
\end{ruledtabular}
\label{tab2}
\end{table}

\begin{table}[t]
\caption{E1, M1 and E2 matrix elements (in au) for the considered transitions in the Mn IX, Fe X, Co XI and Ni XII ions
coming from the DHF, DC, (DC+Breit), (DC+Breit+VP) and (DC+Breit+VP+SE) approximations. We have given these results 
from the CCSD method and contributions from the partial triple excitations are added at the end. The recommended 
values from our calculations with uncertainties are quoted as ``Reco''.}
\begin{ruledtabular}
\begin{tabular}{lcccc}
Method& \multicolumn{2}{c}{$3s^23p^5   \ ^2P_{1/2}\rightarrow $} & $3s3p^6   \ ^2S_{1/2} \rightarrow$ & $3s3p^6   \ ^2S_{1/2} \rightarrow$ \\
      & \multicolumn{2}{c}{$3s^23p^5 \ ^2P_{3/2}$}    & $3s^23p^5   \ ^2P_{1/2}$           & $3s^23p^5   \ ^2P_{3/2}$ \\
\cline{2-3} \\  
 &  M1 & E2 &  E1 & E1  \\
\hline \\
 Mn IX & & \\
DHF        & $1.15421$  & $0.68517$  & 0.65197 & $0.92508$  \\
DC         & $1.15082$  & $0.59500$  & 0.35195 & $0.40368$ \\
$+$Breit   & $1.15079$  & $0.59540$  & 0.35163 & $0.40330$ \\
$+$VP      & $1.15080$  & $0.59540$  & 0.35160 & $0.40324$ \\
$+$SE      & $1.15080$  & $0.59483$  & 0.35211 & $0.40455$   \\
$+$Triples & $1.15078$  & $0.59486$  & 0.35316 & $0.40660$ \\
  & & \\
Reco       & $1.1508(1)$& $0.5948(4)$& 0.353(1)& $0.406(1)$ \\
\hline \\
 Fe X & & \\
DHF        & $1.15414$  & $0.59914$  & 0.61202 & $0.86869$ \\
DC         & $1.15187$  & $0.52127$  & 0.33404 & $0.38622$ \\
$+$Breit   & $1.15185$  & $0.52177$  & 0.33475 & $0.38783$ \\
$+$VP      & $1.15184$  & $0.52172$  & 0.33369 & $0.38576$ \\
$+$SE      & $1.15185$  & $0.52133$  & 0.33438 & $0.38725$   \\ 
$+$Triples & $1.15183$  & $0.52135$  & 0.33529 & $0.38923$ \\
  & & \\
Reco       & $1.1518(1)$&$0.5213(2)$ & 0.335(1)& $0.389(1)$ \\
\hline \\
 Co XI & & \\
DHF        & $1.15406$  & $0.52867$  & 0.57670 & $0.81887$ \\
DC         & $1.15257$  & $0.46292$  & 0.31750 & $0.36755$ \\
$+$Breit   & $1.15254$  & $0.46335$  & 0.31714 & $0.36717$ \\
$+$VP      & $1.15253$  & $0.46336$  & 0.31710 & $0.36711$ \\
$+$SE      & $1.15255$  & $0.46386$  & 0.31768 & $0.36877$   \\ 
$+$Triples & $1.15253$  & $0.46398$  & 0.31866 & $0.37071$ \\
  & & \\
Reco       & $1.1525(1)$& $0.464(1)$& 0.319(1)& $0.370(1)$ \\
\hline \\
 Ni XII & & \\
DHF        & $1.15397$  & $0.47016 $ & 0.54525 & $0.77451$ \\
DC         & $1.15214$  & $0.43969 $ & 0.28383 & $0.34649$ \\
$+$Breit   & $1.15243$  & $0.41536 $ & 0.30295 & $0.35314$ \\
$+$VP      & $1.15243$  & $0.41538 $ & 0.30292 & $0.35307$ \\
$+$SE      & $1.15250$  & $0.41472 $ & 0.30392 & $0.35480$    \\ 
$+$Triples & $1.15250$  & $0.41473$  & 0.30479 & $0.35653$ \\
  & & \\
Reco       & $1.1525(2)$& $0.4147(1)$&0.3048(3)& $0.3565(1)$ \\

\end{tabular}
\end{ruledtabular}
\label{tab3}
\end{table}

\begin{table}[t]
\caption{Contributions to the E1, M1 and E2 amplitudes from various RCC terms of the CCSD(T) method using the (DC+Breit+VP+SE) Hamiltonian in the Mn IX and Fe X ions.}
\begin{ruledtabular}
\begin{tabular}{lcccc}
RCC   & \multicolumn{2}{c}{$3s^23p^5   \ ^2P_{1/2} \rightarrow$} & $3s3p^6   \ ^2S_{1/2} \rightarrow$ & $3s3p^6   \ ^2S_{1/2} \rightarrow$ \\
 term & \multicolumn{2}{c}{$3s^23p^5 \ ^2P_{3/2}$}               & $3s^23p^5   \ ^2P_{1/2}$           & $3s^23p^5   \ ^2P_{3/2}$ \\
\cline{2-3} \\  
 &  M1 & E2 &  E1 & E1  \\
\hline \\
 Mn IX & & \\
$\overline{O}_{ob}$                          &-1.14501 &-0.66667 &0.64321 &-0.91119   \\
$\overline{O}_{ob}R_{1a}$                    &0.00001  &0.00006  &-0.00005& 0.00008   \\
$R_{1a}^{\dagger} \overline{O}$              &0.00001  &0.00006  &0.00006 &-0.00009    \\
$\overline{O}_{ob}R_{2a}$                    &0.00039  &0.03221  &-0.09488&0.13285     \\
$R_{2a}^{\dagger} \overline{O}$              &-0.00030 &0.03211  &-0.22270&0.32237     \\
$R_{1a}^{\dagger}\overline{O}_{ob}R_{1a}$    &$\sim 0$ &$\sim 0$ &$\sim 0$&$\sim 0$    \\
$R_{1a}^{\dagger}\overline{O}_{ob}R_{2a}$    &$\sim 0$ &$\sim 0$ &$\sim 0$&$\sim 0$    \\
$R_{2a}^{\dagger}\overline{O}_{ob}R_{1a}$    &$\sim 0$ &$\sim 0$ &$\sim 0$&$\sim 0$    \\
$R_{2a}^{\dagger}\overline{O}_{ob}R_{2a}$    &-0.01899 &0.00323  &0.02624 &0.05171     \\
Others                                       &-0.00094 &-0.00313 &0.00278 &-0.00400    \\
$norm$                                       &0.01408  &0.00727  &-0.00150&0.00167     \\
\hline \\
 Fe X & & \\
$\overline{O}_{ob}$                          &-1.14547 &-0.58110 &0.60414 &-0.85686   \\
$\overline{O}_{ob}R_{1a}$                    &0.00001  &0.00005  &-0.00004&0.00007     \\
$R_{1a}^{\dagger} \overline{O}$              &-0.00001 &0.00004  &0.00007 &-0.00010    \\
$\overline{O}_{ob}R_{2a}$                    &0.00039  &0.02664  &-0.08733&0.12207     \\
$R_{2a}^{\dagger} \overline{O}$              &-0.00030 &0.02658  &-0.20625&0.30004     \\
$R_{1a}^{\dagger}\overline{O}_{ob}R_{1a}$    &$\sim 0$ &$\sim 0$ &$\sim 0$&$\sim 0$    \\
$R_{1a}^{\dagger}\overline{O}_{ob}R_{2a}$    &$\sim 0$ &$\sim 0$ &$\sim 0$&$\sim 0$   \\
$R_{2a}^{\dagger}\overline{O}_{ob}R_{1a}$    &$\sim 0$ &$\sim 0$ &$\sim 0$&$\sim 0$    \\
$R_{2a}^{\dagger}\overline{O}_{ob}R_{2a}$    &-0.01925 &0.00268  &0.02369 &0.04742    \\
Others                                       &-0.00080 &-0.00237 &0.00237 &-0.00340   \\
$norm$                                       &0.01362  &0.00613  &-0.00136&0.00153    \\
\end{tabular}
\end{ruledtabular}
\label{tab4}
\end{table}

\begin{table}[t]
\caption{Contributions to the E1, M1 and E2 amplitudes from various RCC terms of the CCSD(T) method using the (DC+Breit+VP+SE) Hamiltonian in the Co XI and Ni XII ions.}
\begin{ruledtabular}
\begin{tabular}{lcccc}
RCC   & \multicolumn{2}{c}{$3s^23p^5   \ ^2P_{1/2} \rightarrow$} & $3s3p^6   \ ^2S_{1/2} \rightarrow$ & $3s3p^6   \ ^2S_{1/2} \rightarrow$ \\
 term & \multicolumn{2}{c}{$3s^23p^5 \ ^2P_{3/2}$}               & $3s^23p^5   \ ^2P_{1/2}$           & $3s^23p^5   \ ^2P_{3/2}$ \\
\cline{2-3} \\  
 &  M1 & E2 &  E1 & E1  \\
\hline \\
 Co XI & & \\
$\overline{O}_{ob}$                          &-1.14578 &-0.51431 &0.56952 &-0.80751   \\
$\overline{O}_{ob}R_{1a}$                    &0.00001  &0.00004  &-0.00004&0.00006     \\
$R_{1a}^{\dagger} \overline{O}$              &-0.00001 &0.00004  &0.00007 &-0.00010   \\
$\overline{O}_{ob}R_{2a}$                    &0.00040  &0.02233  &-0.08086&0.11282     \\
$R_{2a}^{\dagger} \overline{O}$              &-0.00030 &0.02232  &-0.19250&0.28160      \\
$R_{1a}^{\dagger}\overline{O}_{ob}R_{1a}$    &$\sim 0$ &$\sim 0$ &$\sim 0$&$\sim 0$    \\
$R_{1a}^{\dagger}\overline{O}_{ob}R_{2a}$    &$\sim 0$ &$\sim 0$ &$\sim 0$&$\sim 0$   \\
$R_{2a}^{\dagger}\overline{O}_{ob}R_{1a}$    &$\sim 0$ &$\sim 0$ &$\sim 0$&$\sim 0$    \\
$R_{2a}^{\dagger}\overline{O}_{ob}R_{2a}$    &-0.01930 &0.00226  &0.02166 &0.04392   \\
Others                                       &-0.00069 &-0.00193 &0.00204 &-0.00290   \\
$norm$                                       &0.01316  &0.00527  &-0.00123&0.00140   \\
\hline \\
 Ni XII & & \\
$\overline{O}_{ob}$                          &-1.14539 &-0.45742 &0.53861 &-0.76408   \\
$\overline{O}_{ob}R_{1a}$                    &0.00001  &0.00003  &-0.00003&0.00005     \\
$R_{1a}^{\dagger} \overline{O}$              &-0.00001 &0.00003  &0.00008 &-0.00011     \\
$\overline{O}_{ob}R_{2a}$                    &0.00041  &0.01880  &-0.07501&0.10445      \\
$R_{2a}^{\dagger} \overline{O}$              &-0.00030 &0.01882  &-0.17924&0.26391      \\
$R_{1a}^{\dagger}\overline{O}_{ob}R_{1a}$    &$\sim 0$ &$\sim 0$ &$\sim 0$&$\sim 0$     \\
$R_{1a}^{\dagger}\overline{O}_{ob}R_{2a}$    &$\sim 0$ &$\sim 0$ &$\sim 0$&$\sim 0$     \\
$R_{2a}^{\dagger}\overline{O}_{ob}R_{1a}$    &$\sim 0$ &$\sim 0$ &$\sim 0$&$\sim 0$    \\
$R_{2a}^{\dagger}\overline{O}_{ob}R_{2a}$    &-0.01917 &0.00191  &0.01975 &0.04050      \\
Others                                       &-0.00069 &-0.00146 &0.00176 &-0.00255     \\
$norm$                                       &0.01274  &0.00456  &-0.00113&0.00130     \\
\end{tabular}
\end{ruledtabular}
\label{tab5}
\end{table}
We take the help of diagrammatic representation to get solutions in an easier way for the above equations. In 
this process, we divide first as $H_N=F_N+V_N$ with $F_N$ representing the DHF 
Hamiltonian which is an effective one-body operator and $V_N$ is the normal ordering form of the residual Coulomb-Breit 
interaction. By construction, $F_N$ is diagonal in nature and the one-body contributions from $V_N$ cancels out in
our calculations. Following the Koopman's theorem \cite{Lindgren}, the detachment energy of an electron from orbital $a$ 
at the DHF level is just the diagonal value of the operator $F_N$ (the single particle orbital energy of $a$). 
To minimize the computational time, we construct effective one-body and two-body intermediate terms from $\widehat{F_N e^T}$ and 
$\widehat{V_N e^T}$ as shown in Figs. \ref{fig2} and \ref{fig3}, respectively, and connect them finally with the $R_a$ operators to 
solve the above energy and amplitude equations. The energy evaluating diagrams are shown in Fig. \ref{fig4}.
Similarly, the diagrams contributing to the $R_{1a}$ and $R_{2a}$ amplitude calculations are shown in Figs. \ref{fig5} and
\ref{fig6}, respectively.

The quality of the results are further elevated with the consideration of the most important triple excitation
configurations from $|\Phi_a \rangle$ by constructing a perturbative RCC operator $R_{3a}^{pert}$ as
\begin{eqnarray} 
R_{3a}^{pert} = \frac{1}{12} \sum_{pr,bdg} \frac{(\widehat{H_NT_2}+\widehat{H_N R_{2a}})^{prb}_{dga}}
{(\varepsilon_b+\varepsilon_d+\varepsilon_g-\varepsilon_a-\varepsilon_p-\varepsilon_r)}, 
\end{eqnarray}
with $\varepsilon$s representing the single particle orbital energies. Instead of considering this operator explicitly,
we account its contributions implicitly in the self-consistent evaluation of $\Delta E_a$. This approach is
usually referred to as CCSD(T) method in the literature. Diagrammatic representation of the $R_{3a}^{pert}$
operator are given in Fig. \ref{fig7}.

We evaluate $\Delta E_a$s using the laboratory value as $c=137.03599972$ and modify the $c$ value suitably for the corresponding 
$x$ values to obtain the transition frequencies, $\omega(x)$, between all possible states that are of our interest. 

Once any two given $|\Psi_f \rangle$ and $|\Psi_i \rangle$ states are obtained in the above procedure, the matrix element
of an operator $O$ between these two states are evaluated using the expression
\begin{eqnarray}
\frac{\langle \Psi_f | O | \Psi_i \rangle}{\sqrt{\langle \Psi_f|\Psi_f\rangle \langle \Psi_i|\Psi_i\rangle}} &=& \frac{\langle \Phi_f | \{ 1+ R_f^{\dagger}\} \overline{O}
\{ 1+ R_i\} |\Phi_i\rangle}{ \sqrt{ {\cal{N}}_f {\cal{N}}_i  }}, \nonumber \\
\label{eqn25}
\end{eqnarray}
where $\overline{O}=(e^{T^{\dagger}} O e^T)_l$ and ${\cal N}_i = \{ (1+R_i^{\dagger}) \overline{\cal{N}} (1+R_i) \}$ with $\overline{\cal N} = (e^{T^{\dagger}} e^T)_l$, 
for the subscript $l$ means only the linked terms are the contributing terms, involves two non-truncative series in the above
expression whose contributions are accounted as much as possible in stepwise. To do so, we divide $\overline{O}$ and $\overline{\cal N}$ into
the effective fully contracted, one-body, two-body etc. terms in the diagrammatic form by employing the generalized
Wick's theorem \cite{Lindgren}. Since these terms are either connected with the $R_a$ operators or has to be 
the effective one-body term for the consideration at the final stage property calculation, as a result they get truncated 
factitiously at the effective five-body terms in the CCSD method. The intermediate storage of the effective 
three-body terms onwards is an affair of huge computational cost and direct calculation of these diagrams will 
be enormously time consuming against a very little contribution to the final result. Therefore, we have
neglected these contributions on the basis that they belong to the class of diagrams with fifth or higher orders in the residual 
Coulomb-Breit interaction. We first calculate the intermediate effective one-body diagrams of hole-hole (H-H),
particle-particle (P-P), hole-particle (H-P) and particle-hole (P-H) types from $\overline{O}$ and $\overline{\cal N}$
considering terms up to minimum fifth order in the residual Coulomb-Breit interaction and store these intermediate parts
for their further use. It has been found in our study, as will be demonstrated in the next Results and Discussion 
section, that the H-P and P-H diagrams carrying out the core-polarization effects to all orders are contributing 
predominantly in the considered ions. Therefore, we have replaced the corresponding $O$ operator from the P-H and H-P 
effective diagrams by the P-P and H-H diagrams to dress-up further the effective H-P/P-H operators for accounting
these contributions as rigorously as possible. All these four types of effective one-body terms are then connected with the $R_a$ 
and its complex-conjugate ($cc$) diagrams to obtain results for the final calculation. These final contributing 
diagrams are shown in Fig. \ref{fig8}.

We then formulate the effective two-body terms from $\overline{O}$ and $\overline{\cal N}$ in the following way to
account their contributions at the minimum computational requirements. We connect the effective one-body terms of 
$\overline{O}$ with another $T$ and with its $cc$ operators to form the effective two-body diagrams. This procedure
obviously takes into account more higher order terms than the two-body terms that could have been generated by connecting only the operator $O$ with 
the $T$ operators. Unlike the effective one-body terms, effective two-body terms are computed directly after contracting with the $R_a$ 
operators. Some of the important effective two-body diagrams contributing substantially in the present calculations are shown 
in Fig. \ref{fig9}.

Contribution to the matrix element after the normalizations of the wave functions ($norm$) is estimated explicitly using the
expression
\begin{eqnarray}
norm &=& \left [ \frac{\langle \Psi_f | O | \Psi_i \rangle} {\sqrt{\langle \Psi_f|\Psi_f\rangle
\langle \Psi_i|\Psi_i\rangle}} - \langle \Psi_f | O | \Psi_i \rangle \right ] \nonumber \\
  &=& \left [ \frac{1}{ \sqrt{ {\cal{N}}_f {\cal{N}}_i  }} - 1 \right ] \langle \Psi_f | O | \Psi_i \rangle .
\label{eqn26}
\end{eqnarray}

\begin{table}[t]
\caption{Transition rates ($A$ in $s^{-1}$) and oscillator strengths ($f$) in the considered ions.}
\begin{ruledtabular}
\begin{tabular}{lccc}
Transition &  \multicolumn{2}{c}{$A_{k i}$} & $f_{k i}$ \\
\cline{2-3} \\  
 &  Others & Present &   \\
 &    $^a$\cite{Kaufman},$^b$\cite{Brenner} &  &   \\
\hline \\
 Mn IX & & \\
$3s^23p^5   \ ^2P_{1/2} \xrightarrow{M1} 3s^23p^5   \ ^2P_{3/2}$ & $^a$35.5      &35.27(1)    &1.67(1)[-7]   \\
       
$3s^23p^5   \ ^2P_{1/2} \xrightarrow{E2} 3s^23p^5   \ ^2P_{3/2}$ &          &6.16(1)[-3] &2.93(1)[-11]  \\ 

$3s3p^6     \ ^2S_{1/2} \xrightarrow{E1} 3s^23p^5   \ ^2P_{1/2}$ &          &2.04(1)[9]  &4.76(3)[-2]   \\
       
$3s3p^6     \ ^2S_{1/2} \xrightarrow{E1} 3s^23p^5   \ ^2P_{3/2}$ &          &3.13(1)[9]  &3.31(2)[-2]   \\
\hline \\
 Fe X & & \\
$3s^23p^5   \ ^2P_{1/2} \xrightarrow{M1} 3s^23p^5   \ ^2P_{3/2}$  & $^a$69.4     &69.01(1)   &2.10(2)[-7]    \\
                                                                  & $^b$70.4 & & \\
       
$3s^23p^5   \ ^2P_{1/2} \xrightarrow{E2}  3s^23p^5   \ ^2P_{3/2}$  &         &1.44(1)[-2] &4.41(1)[-11]   \\

$3s3p^6     \ ^2S_{1/2} \xrightarrow{E1}  3s^23p^5   \ ^2P_{1/2}$  &         &2.33(1)[9]  &4.64(3)[-2]   \\
       
$3s3p^6     \ ^2S_{1/2} \xrightarrow{E1}  3s^23p^5   \ ^2P_{3/2}$  &         &3.71(2)[9]  &3.31(2)[-2]    \\
\hline \\
Co XI & & \\

$3s^23p^5   \ ^2P_{1/2} \xrightarrow{M1}  3s^23p^5   \ ^2P_{3/2}$  & $^a$130.0    &129.68(2)   &2.60(1)[-7]  \\ 
       
$3s^23p^5   \ ^2P_{1/2} \xrightarrow{E2}  3s^23p^5   \ ^2P_{3/2}$  &         &3.26(1)[-2] &6.54(1)[-11] \\

$3s3p^6     \ ^2S_{1/2} \xrightarrow{E1}  3s^23p^5   \ ^2P_{1/2}$  &         &2.62(2)[9]  &4.51(3)[-2]   \\ 
       
$3s3p^6     \ ^2S_{1/2} \xrightarrow{E1}  3s^23p^5   \ ^2P_{3/2}$  &         &4.29(2)[9]  &3.25(2)[-2]  \\
\hline \\
 Ni XII & & \\
$3s^23p^5   \ ^2P_{1/2} \xrightarrow{M1}  3s^23p^5   \ ^2P_{3/2}$  & $^a$237.0    &236.31(3)   &3.17(1)[-7]    \\
       
$3s^23p^5   \ ^2P_{1/2} \xrightarrow{E2}  3s^23p^5   \ ^2P_{3/2}$  &         &7.10(1)[-2] &9.53(1)[-11]    \\

$3s3p^6     \ ^2S_{1/2} \xrightarrow{E1}  3s^23p^5   \ ^2P_{1/2}$  &         &2.94(1)[9]  &4.42(1)[-2]    \\
       
$3s3p^6     \ ^2S_{1/2} \xrightarrow{E1}  3s^23p^5   \ ^2P_{3/2}$  &         &4.99(2)[9]  &3.24(1)[-2]     \\

\end{tabular}
\end{ruledtabular}
\label{tab6}
\end{table}

\begin{table}[t]
\caption{Lifetimes ($\tau$) of the first two excited states (in $ms$) of the considered ions.}
\begin{ruledtabular}
\begin{tabular}{lcccc}
State & This work & Other & Experiment \\
      &           & prediction & \\
\hline \\
Mn IX & & \\

        $3s^23p^5 \ ^3P_{1/2}$  &$28.34(2)$    &              &                             \\

        $3s3p^6   \ ^2S_{1/2}$  &$1.93(3)[-7]$ &                       &                              \\
\hline \\
Fe X & & \\
        $3s^23p^5 \ ^3P_{1/2}$  &$14.48(2)$    &14.46$^a$, 14.41$^b$   &13.64($\pm$0.25)$^k$           \\
                                &              &14.41$^c$              &14.41($\pm$0.14)$^l$            \\
                                &              &14.40$^d$, 14.39$^e$   & 14.2($\pm$0.2)$^m$                                \\ 
                                &              &14.46$^f$, 14.37$^g$   &                                 \\  
                                &              &16.60$^h$, 18.21$^i$   &                                  \\
                                &              &15.29$^j$, 14.42$^k$   &                            \\
                                &              &                       &                           \\    

        $3s3p^6   \ ^2S_{1/2}$  &$1.65(3)[-7]$ &                       &                             \\
\hline  \\
Co XI & & \\
        $3s^23p^5 \ ^3P_{1/2}$  &$7.71(2)$     &7.69$^b$, 7.69$^c$     &7.62($\pm$0.46)$^l$             \\ 
                                &              & 8.67$^h$              &                                \\  

         $3s3p^6   \ ^2S_{1/2}$ &$1.44(3)[-7]$&                        &                                 \\
\hline \\
Ni XII & & \\
        $3s^23p^5 \ ^3P_{1/2}$  &$4.23(2)$     &4.22$^{b,c,d,g}$       &4.166($\pm$0.06)$^l$             \\
                                &              &4.69$^h $    &                                \\ 

        $3s3p^6   \ ^2S_{1/2}$  &$1.26(3)[-7]$ &                       &                                \\
\end{tabular}
\end{ruledtabular}
\begin{tabular}{lc}
References: & $^a$ \cite{Krueger}. \\ 
            & $^b$ \cite{Warner}. \\  
            & $^c$ \cite{Smith}. \\
            & $^d$ \cite{Kastner}.  \\
            & $^e$ \cite{Mason}.\\
            & $^f$ \cite{Kafatos}. \\
            & $^g$ \cite{Eidelsberg}.\\
            & $^h$ \cite{Huang}. \\   
            & $^i$ \cite{Bhatia}. \\
            & $^j$ \cite{Kohstall, Dong}. \\ 
            & $^k$ \cite{Moehs1, Moehs2}.\\
            & $^l$ \cite{Trabert3}.\\
            & $^m$ \cite{Brenner}.\\
\end{tabular}   
\label{tab7}        
\end{table}

\section{Results and Discussion}

We present the detachment energies obtained using our above described methods 
in Table \ref{tab1} at the various levels of approximations in the Hamiltonian
for all the considered ions. We give contributions from DHF and CCSD(T) results 
with gradual changes in the calculated values 
after the inclusion of Breit, VP and SE interactions. We also give the 
differences between the CCSD(T) and CCSD results within the parentheses of the given
CCSD(T) results to demonstrate the importance of including the triple
excitation configurations. It can also be noticed here that the CCSD(T) method
improves the results over the CCSD approach in all the states.
Our results are also compared with experimental 
values listed in the national institute of science and technology (NIST) 
database \cite{Nist}. As can be seen, contributions from the higher order
relativistic corrections are not small in the evaluation of these quantities.
Among them the SE interaction are the largest contributing relativistic
corrections. The differences between our final results with the full Hamiltonian
and NIST results are given as $\delta$ in the same table which shows that the
CCSD(T) results are sub-one percent accurate for each state in all the four ions.
We also observe that the ratios $y=(\Delta E_a^{DC} - \Delta E_a^{final})/\Delta E_a^{final}$, 
with $\Delta E_a^{DC}$ is the contribution from the DC Hamiltonian and $\Delta E_a^{final}$
is the final result, are almost same in all the states except in the excited 
$3s^23p^5 \ ^2P_{1/2}$ state where it is slightly large. These values are comparatively
larger in the Ni XII ion implying that the relativistic effects are increasing with 
the size of the ion.

The accuracies attained in the energy calculations for the considered ions seem to be very promising to investigate
the relativistic dependency in these quantities for the study of possible variation of $\alpha_e$ 
by determining the sensitivity coefficients $q$ of the transitions among the calculated states. These
coefficients are given in Table \ref{tab2} with the DC Hamiltonian and with other relativistic corrections
using the CCSD(T) method.
The obtained results are quite enhanced in these ions and the values increase for the heavier 
ions. We also observe that the corrections due to the Breit and QED interactions are influencing the results
considerably which are never investigated before in the other studied highly charged ions \cite{berengut1,
berengut2}. It can be found that the contributions from the QED corrections are almost
negligible in the $\Delta E_a$ calculations, however these contributions are found
to be relatively large in the determination of $q$ values. Since our calculated $\Delta E_a$ values are below 
0.5 \% accurate compared with their experimental values, on this ground we recommend that these reported 
$q$ values are also accurate within the same percentage.

We now turn to determining other properties of the transitions whose sensitivity coefficients are estimated in this work. The important
transition properties that should be known precisely for their astrophysical observations are the transition probabilities, the oscillator
strengths and the lifetimes of the considered excited states. The transitions from the fine structure level to the ground state in these
ions decay through the M1 and E2 forbidden channels while the $3s3p^6 \ ^2S_{1/2}$ state decay to the $3s^23p^5 \ ^2P_{1/2}$ state and
to the ground state via the E1 channel. The transition amplitudes for these channels obtained from our calculations with different
approximations are given in Table \ref{tab3}. At the end, we also give the recommended values as ``Reco'' 
with maximum probable uncertainties associated with these values. These uncertainties are estimated based on the 
intuitive guess from the trends they exhibit using the CCSD and CCSD(T) methods and at various
approximations in the Hamiltonian. It can also be noticed that the DHF results are large from the RCC calculations and the differences between the 
DHF and RCC results are small for the M1 transition amplitudes which are large in the E2 amplitudes and the RCC results are almost 
half of the DHF results in the E1 amplitudes persuading large correlation effects in this property. 
To understand the role of various
correlation effects in the RCC calculations of these quantities, we give contributions explicitly 
from various terms of the CCSD(T)
method using the full Hamiltonian in Tables \ref{tab4} and \ref{tab5}. As can be seen, the effective one-body contribution through $\overline{O}$
involving the DHF result is the most dominant contributing term followed by the effective two-body terms for the M1 transition else
the $\overline{O}_{ob}R_{2a}$ term along with its $cc$ term in the E1 and E2 amplitude calculations. The reason for $\overline{O}_{ob}R_{2a}$ and
effective two-body contributions being very large as they account directly the core-polarization contributions
to all orders involving the valence electrons which are found to be very crucial in the considered ions. Nevertheless, the contributions
from $norm$ are non-negligible.

Using the above transition amplitudes, we give the transition probabilities and the oscillator strengths 
for the considered ions in Table \ref{tab6}. To estimate these quantities, we have used the experimental
energies to avoid the uncertainties coming out from the calculated energies although these calculations are
sufficiently accurate to provide precise {\it ab initio} values. There are estimation of the transition 
probabilities due to the M1 transitions earlier \cite{Kaufman} which were determined using the M1 
amplitudes obtained using the DHF method and experimental energies. Since correlation effects are very small
in the calculation of the M1 amplitudes in the considered transitions, we see a very good agreement between
both the work. Recently, the transition probability of the $3s^23p^5  \ ^2P_{1/2} \rightarrow 3s^23p^5 \ ^2P_{3/2}$
transition of the Fe X ion is measured by Brenner {\it et al.} \cite{Brenner} which also agree with our result. But our 
result seem to be more precise than these two reported values. The oscillator strengths for the allowed
transitions are found to be large enough to be used for the detection of these lines in the 
astrophysical observations.

Finally, we present the lifetimes of the excited states in Table \ref{tab7}. These values are
compared with the previously reported experimental and predicted values. As seen in the table,
the experimental values have large uncertainties except for the $3s^23p^5 \ ^2P_{1/2}$ state of 
Ni XII. The lifetime of the $3s^23p^5 \ ^2P_{1/2}$ state of Fe X was measured in a Kingdon ion trap 
\cite{Moehs1, Moehs2} which differs from other measurements that are carried out 
optically in a heavy-ion storage ring \cite{Trabert3} and using the electron
beam ion-trap technique \cite{Brenner}. Our result agrees with the latter two
measurements. The lifetimes for the same state in 
Co XI and Ni XII are also measured by Tr\"abert and coworkers \cite{Trabert3} which agree with our 
estimated values, however our theoretical values seem to be more precise than the measurements. The other
predicted values of the lifetime of this state in all these three ions \cite{Krueger,Warner, Smith, Kastner,
Mason, Kafatos, Eidelsberg, Huang, Kohstall, Dong} are either obtained from the astrophysical observations or 
estimated using lower order many-body methods than our RCC method. We could not find out any reported values for the lifetimes 
of the above two excited states of Mn IX in the literature. Our estimated results for the
lifetimes of the excited states in this ion will be useful for their measurements.

\section{Conclusion}

We have developed a relativistic coupled-cluster method to calculate atomic wave functions of the states in ions
which have one electron less than the closed-shell electronic configurations. We successfully employed this
theory to calculate the wave functions in the highly charged Mn IX, Fe X, Co XI and Ni XII ions. The Dirac-Coulomb
Hamiltonian with other relativistic interactions such as Breit, vacuum polarization and self energy corrections
is used to incorporate both the relativistic and correlation effects more rigorously in the calculations.
Configuration interaction space is approximated at the singles and doubles excitation level, however they
are elevated by the inclusion of the important triple excitations in a self-consistent manner through 
a perturbative approach. We have obtained the detachment energies within sub-one percent accuracy and 
estimated the sensitivity coefficients for the investigation of any possible temporal variation
of the fine structure constant at the same level of accuracy. Roles of various relativistic and correlation
effects are demonstrated explicitly. Further more, we determined the transition matrix elements due to
the E1, M1 and E2 channels from the considered excited states in the above ions. Using these matrix elements,
we evaluated the transition probabilities, the oscillator strengths and the lifetimes of the excited states 
and compared them with the available experimental and other predicted values. Our estimated results are found
to be more precise than the previously reported results. The corresponding experimental results for some of 
our reported values are not known, hence our calculated values will serve as the benchmark results for their 
future measurements.

\section*{Acknowledgment}
We thank Yashpal Singh for his partial contributions in the developed method. The calculations were carried out using 
PRL 3TFLOP HPC cluster, Ahmedabad.

\end{document}